\documentclass[aps,prl,twocolumn,superscriptaddress,showpacs]{revtex4-1}
\usepackage[pdftex]{graphicx}
\usepackage{color}

\begin{document}


\title{Assisted diffusion of buried hydrogen in Pd by means of
inelastic ballistic electrons}
\title{Ballistic-electrons assisted-diffusion of buried hydrogen in Pd}
\title{Diffusion of hydrogen in Pd assisted by inelastic ballistic hot electrons}



\newcommand{\DIPC}[0]{{
Donostia International Physics Center, Universidad del Pa\'{\i}s Vasco UPV/EHU,
Paseo Manuel de Lardiz\'abal 4, 20018 Donostia-San Sebasti\'an, Spain}}
\newcommand{\CFM}[0]{{
Centro de F\'{\i}sica de Materiales CFM/MPC (CSIC-UPV/EHU), Paseo Manuel de
Lardiz\'abal 5, 20018 Donostia-San Sebasti\'an, Spain}}
\newcommand{\QUIM}[0]{{
Departamento de F\'{\i}sica de Materiales, Facultad de Qu\'{\i}micas UPV/EHU,
Apartado 1072, 20018 Donostia-San Sebasti\'an, Spain}}
\newcommand{\ICMM}[0]{{
Instituto de Ciencia de Materiales de Madrid, CSIC,
Cantoblanco, 28049 Madrid, Spain}}

\author{M. Blanco-Rey}
\email[]{maria\_blancorey@ehu.es}
\affiliation{\DIPC}

\author{M. Alducin}
\affiliation{\CFM}

\author{J.I. Juaristi}
\affiliation{\CFM}
\affiliation{\QUIM}

\author{P.L. de Andres}
\affiliation{\DIPC}
\affiliation{\ICMM}


\date{\today}

\begin{abstract}
Sykes {\it et al.} [Proc. Natl. Acad. Sci. {\bf 102}, 17907 (2005)]
have reported how electrons injected from
a scanning tunneling microscope modify the diffusion rates of H buried beneath Pd(111).
A key point in that experiment is the symmetry between positive and negative
voltages for H extraction, which is difficult to explain in view of the large
asymmetry in Pd between the electron and hole densities of states. Combining
concepts from the theory of ballistic electron microscopy and electron-phonon
scattering we show that H diffusion is driven by the
$s$-band electrons only, which explains the observed symmetry.
\end{abstract}


\maketitle

While the scanning tunneling microscope (STM) seems a
natural tool to manipulate all sort of atoms and molecules adsorbed
on surfaces~\cite{bib:stroscio91,bib:stipe98,bib:pascual03,bib:komeda05},
it is rather surprising to learn that tunneling currents can
also be used to change diffusion rates of atoms deeply absorbed
in metals. In an elegant experiment performed on H buried below
the Pd(111) surface, Sykes {\it et al.} have unequivocally proved
that this effect does exist~\cite{bib:sykes05}.
Indeed, the STM has been used to investigate buried
interfaces \cite{bib:prietsch95} or even individual molecules \cite{bib:moeller07},
but the pioneering experiment of Sykes {\it et al.} confirms the STM as a
powerful nanometric technique to control diffusion and reaction
processes at the atomic level not only on surfaces but also inside
metals. The controlled insertion and extraction of H in Pd is
an added interest of this particular experiment because of its
possible applications in H
storage~\cite{bib:schlap01,bib:pundt06,bib:tremblay09}
and in heterogenous catalysis. In this respect, Pd is a recurring
catalyst in hydrocarbon synthesis reactions, particularly in
-C=C- bond breaking, and the reason behind such exceptional
catalycity is precisely linked to the existence of weakly bound
H at subsurface sites~\cite{bib:doyle03}.

\begin{figure}
\includegraphics[scale=0.3]{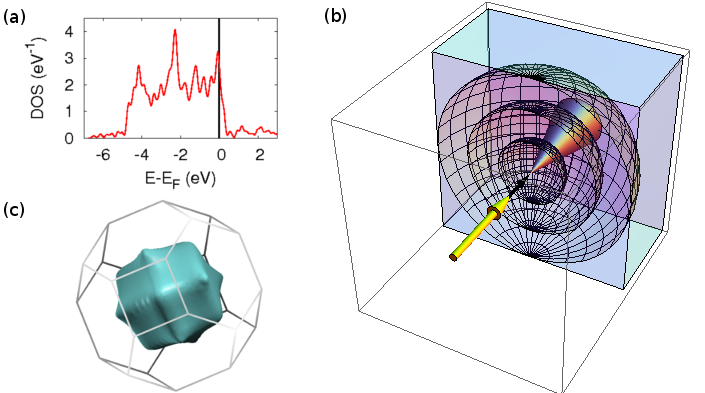}
\caption{\label{fig:esquema}  (a) Calculated density of states
for {\it bulk} Pd. (b) Sketch for injection of $s$-like (isotropic
propagation), and $d$-like (narrowly focused) electrons on Pd(111).
(c) Energy isosurface ($E_{F}+0.83$\,eV) related to the $s$ band
in Pd.}
\end{figure}

Still, the underlying mechanisms behind the selective H population
of subsurface sites reported in~\cite{bib:sykes05} are not well
understood. In particular, the observed bias voltage dependence is
puzzling.  Working at low temperature to prevent thermal diffusion
($T=4$\, K), the STM is used as a local electron gun to inject
small currents ($I_{t} \sim 1-150$\,pA) in the vicinity of the Fermi
energy ($V \sim \pm 1$\,V).  Inelastic excitation with the injected
ballistic hot electrons promotes the diffusion of H from bulk
Pd towards the surface even under low concentration conditions,
thereby implying the contribution of large volumes of the metal
where electric fields are efficiently screened out and cannot play
a role. The distortions in the Pd-Pd distances caused by subsurface
H accumulation are then imaged with STM as protrusions or bright
stripes.  The brightness and width of the stripes increase with $I_t$
and $V$ magnitude. The intriguing finding is that the stripes are
comparable in brightness for opposite $V$ signs, positive
biases creating only slightly wider stripes.  This polarity effect
must have a non trivial explanation, since for $V>0$ and $V<0$
charge carriers are electrons and holes respectively, and the
density of states above and below the Fermi level $E_F$ is very
different in Pd [Fig.~\ref{fig:esquema}~(a)].

In the present letter, we examine the vibrational excitation of H
in Pd caused by inelastic interaction with ballistic hot electrons
using first order perturbation theory for the electron-phonon
(e-ph) coupling. Hopping rates for H migration have been quantified
as a function of the experimental $V$ and $I_t$. Making use of the
theory developed to describe propagation in ballistic electron energy
microscopy (BEEM)~\cite{bib:deandres01}, we find that the observed
symmetry between positive and negative $V$ follows when the different roles of
$s$ and $d$-band electrons has been recognized in the H vibrational
excitation. It has been noticed that $d$-like electrons in {\it bulk} Pd
propagate along narrow cones corresponding to a few selected $k$ points in
the first Brillouin zone (1BZ) \cite{bib:lads03}.
Such a focusing effect happens typically when, with increasing energy,
bands approaching the borders of the 1BZ change their
curvature from convex to concave in order to touch the boundaries at right angles.
While in a typical BEEM experiment focused beams are used to
improve the resolution well below the surface, these electrons
span a reduced substrate volume and have a small probability to
interact with interstitial H in the low dilution regime. On the
contrary, electrons propagating through $s$ channels span large
volumes and are relevant for the present experiment because they
are the only ones that can interact with many interstitial H
[Fig.~\ref{fig:esquema}~(b) and (c)]. Interestingly our result
shows that the inelastic excitation of H in Pd yields information
about a component of the current that is hidden in a typical BEEM
experiment, since it distributes over large spatial regions
and does not contribute to nanometric resolution.
Furthermore, we find that at $T=4$\,K
the efficiency of the diffusion mechanism can only be understood
by incorporating quantum tunneling of H near the top of the barrier.
Similar phenomena have been in fact observed for heavier elements,
e.g., deuterium in ~\cite{bib:Kurland10}.  For this particular case,
a larger mass would reduce the tunneling rate, change the frequencies of
vibrational modes, and modify the rates in general.
However, the relevant physics shall remain the same.

\begin{figure}
\includegraphics[scale=0.7]{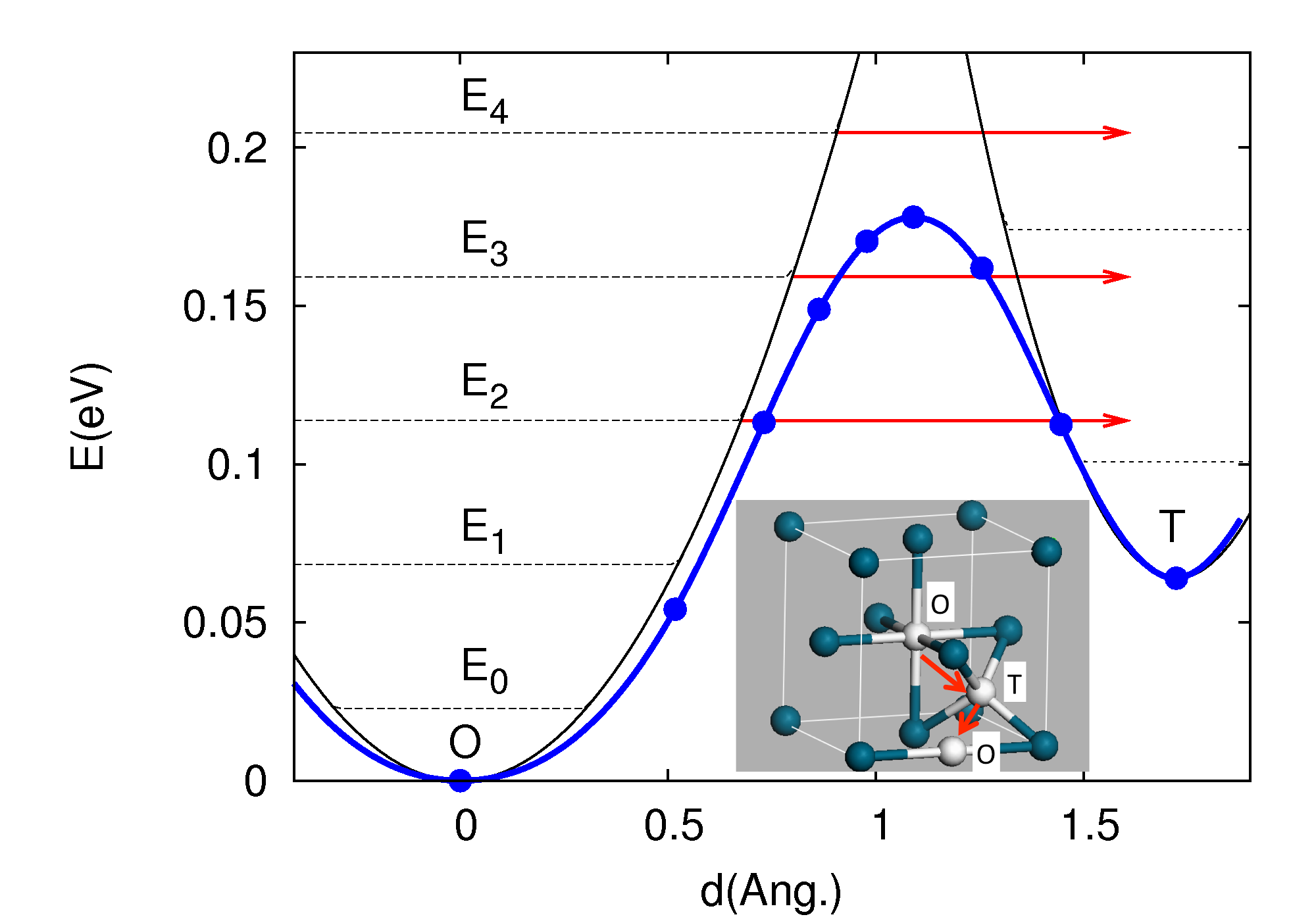}
\caption{\label{fig:barrier} (Color online) Profile of the computed
energy barrier (blue thick line and circles) for diffusion from
the octahedral (O) to the tetrahedral site (T). The approximated
harmonic potential in O and T (black thin lines) and energy
levels (dashed lines) are given for comparison. When H is excited
up to a vibrational level $n \geq 2$, transmission by quantum
tunneling becomes efficient and dominates in the intensity range
of 1--150~pA.
Inset: sketch of O and T sites (white) in the Pd fcc lattice (dark).
Arrows mark the diffusion path from O to T. }
\end{figure}

{\it Ab-initio} calculations have been performed by expanding wave functions on
a plane-wave basis set using the Quantum Espresso package~\cite{bib:qe}.
This formalism yields accurate total energies, equilibrium geometries, electron
band structures and transition states, as well as phonon frequencies,
eigenvectors (in first order density functional perturbation theory), and e-ph
coupling matrix elements. These are the ingredients for the inelastic model
described below. The actual experiment is mimicked by using a periodic
supercell with Pd$_{16}$H stoichiometry. We have checked that such a supercell
brings only small correlations between H atoms in consecutive cells. The theory
level for the exchange and correlation functional is the generalized gradient
approximation in the Perdew-Burke-Ernzerhof formulation~\cite{bib:pbe}, and ion
cores are described by ultrasoft pseudopotentials~\cite{bib:vanderbilt90}. In
the construction of the plane wave basis sets we have used Monkhorst-Pack
$k$-point meshes~\cite{bib:monk76} and energy cut-offs for wave functions and
charge densities of 28 and 180\,Ry, respectively. The convergence threshold for
total energy self-consistency is $10^{-10}$\,Ry. The equilibrium lattice
parameter obtained for fcc Pd is $3.98$\,{\AA}.  Using a $14\times14\times14$
Monkhorst-Pack mesh and the Pd$_{16}$H cell, we determine that H is more stable at
the octahedral site (O) than at the tetrahedral site (T) by 0.064\,eV.  The
atomic positions of the Pd atoms surrounding the H atom have been relaxed with
tolerances of $10^{-8}$\,Ry and $10^{-5}$\,Ry/a.u. in the energies and forces,
respectively. The energy barrier from O to the metastable T has been found by
the nudge elastic band method \cite{bib:henkel00a} to be $E_b =
0.178$\,eV (see Fig.~\ref{fig:barrier}). This is in good agreement with the
0.23\,eV barrier found in electromigration experiments~\cite{bib:pietrzak99}.
In the nudge elastic band calculation, nine images have been used and the final forces at the
transition state are $\lesssim 5\times10^{-4}$\,Ry/a.u.

Phonons have been calculated at the $\Gamma$ point using
$10^{-12}$\,Ry as convergence criterion in the self-consistent
loop. By restricting ourselves to $\vec q = 0$ modes, we are
neglecting correlations between H atoms in consecutive unit cells,
which is exact in the dilute limit.
Only the H atom and its six nearest neighboring Pd atoms are allowed to
move in the phonon calculations. The three nondegenerate
vibrational mode eigenvectors that yield a significant displacement
of the H atom have energies of 17, 21 and 47\,meV.
The constraint on Pd atoms affects these values by only $\sim 5$\,meV.

In order to calculate H diffusion from O to T sites we
approximate the O well by a harmonic potential truncated at an
energy equal to $E_b$~\cite{bib:walkup93,bib:gao97,bib:stipe97}.
Figure~\ref{fig:barrier} shows that the highest frequency mode
$\hbar \omega = 47$\,meV, in good agreement with the barrier profile,
is the most likely to be involved in this process.
Therefore, the H atom ought to be vibrationally excited up
to approximately the fourth harmonic level to overcome the
barrier and diffuse.  The excited level populations obey a
master equation that, in the limit of weak inelasticity, yields
the following transfer rate from the $n$ th vibrational excited
level~\cite{bib:gao97}: \begin{equation} R_n = n \Gamma^{\uparrow}
\bigg( \frac{\Gamma^{\uparrow}}{\Gamma^{\downarrow}} \bigg)^{n-1}
\label{eq:transfer} \end{equation} where $\Gamma^{\uparrow}$
and $\Gamma^{\downarrow}$ are the vibrational excitation and
deexcitation rates between the ground and first excited states.
Hence, $R_{n}$ is obtained as the population of level $n-1$ times
the probability to reach the $n$th vibrational excited state that is
already above the top of the barrier.

Hot electrons injected from the STM tip into the substrate, or in
the inverse direction, have energies in excess according to the bias
voltage, $V$, and can excite vibrational quanta ($\hbar \omega$)
on the interstitial H via e-ph coupling.
In first order time-dependent perturbation theory the transition
rates for electron ($V>0$) or hole ($V<0$) decay can be obtained as:
\begin{eqnarray}
\Gamma^{\uparrow,\downarrow}_1 (\vec q; \omega)& =&
 \frac{2 \pi}{\hbar}
 \frac{1}{N_{\mathrm{Pd}}}
 \sum_{\epsilon_i,\epsilon_j}
 \sum_{\vec k \in \mathrm{1BZ}} \frac{ w^{(s)}_{i \vec k} }{N_{\vec k}}
 | M_{ij}(\vec k, \vec q; \omega) |^2 \nonumber \\
 & & \times \; \delta( \epsilon_j - (\epsilon_i \mp \hbar \omega) )
\label{eq:gamma1}
\end{eqnarray}
\noindent
where $\vec k$ and $\vec q$ are the electron and phonon wavevectors,
respectively.  The $\uparrow,\downarrow$ symbols account for $n
\rightarrow n+1$ (vibrational excitation) and $n \rightarrow n-1$
(vibrational deexcitation) transitions, respectively. The initial
and final electron band energies $\epsilon_{i,j}$ lie in the range
$E_F \leq \epsilon_{i,j} \leq E_F + V$ for bias voltages $V>0$,
and $ E_F-|V| \leq \epsilon_{i,j} \leq E_F$ for $V<0$.
The e-ph coupling matrix elements $M_{ij}(\vec k,\vec q; \omega)$
are computed by Quantum Espresso~\cite{bib:qe}. Convergence on the sum over the
1BZ in Eq.~\ref{eq:gamma1} has been checked
by using a sufficiently fine $k$-point grid~\footnote{Summation in
Eq.~\ref{eq:gamma1} collects contributions from small regions on the
1BZ; to perform the summation accurately the $M_{ij}({\vec k},0;
\omega)$ matrix elements are interpolated on a three times finer
$k$-point grid than the original $14\times14\times14$ mesh. Errors
in $\Gamma_1$ brought by 1BZ sampling effects are hence reduced to
$\lesssim 0.7$\,GHz. The Dirac $\delta$ function is numerically
calculated as a smearing Lorentzian function with half-width $s
\ll \hbar \omega$. The choice $s=6$\,meV yields errors $\lesssim
0.3$\,GHz in $\Gamma_1$.}. $N_{\mathrm{Pd}}$ is the number of Pd
atoms in the unit cell, a normalization factor ensuring that the
rates do not depend on the cell size.

\begin{figure}
\includegraphics[scale=0.7]{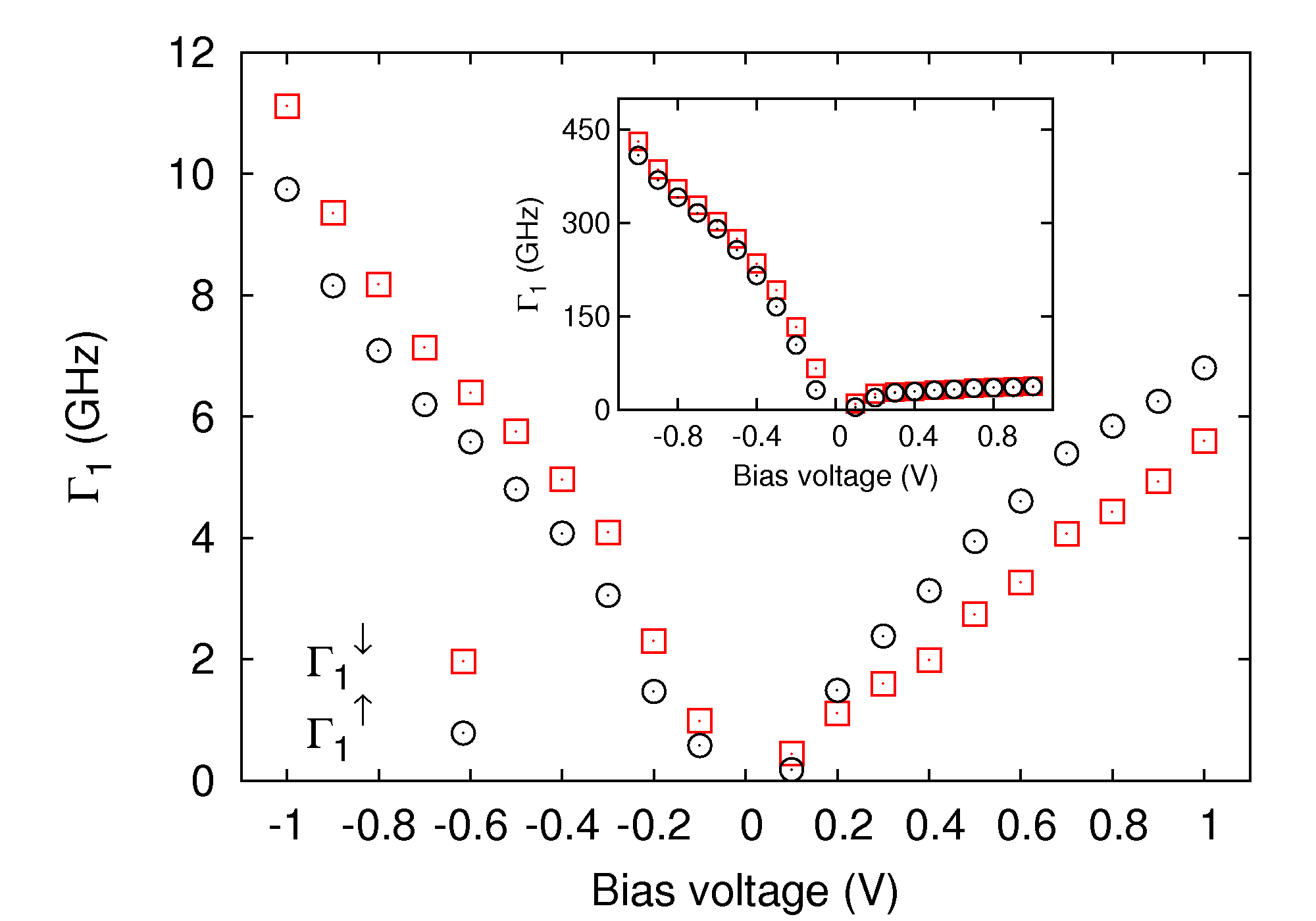}
\caption{\label{fig:gamma1} Bias voltage dependence of the vibrational
excitation ($\Gamma^{\uparrow}_1$) and deexcitation ($\Gamma^{\downarrow}_1$)
rates as calculated from Eq.~(\ref{eq:gamma1}) when including the propagation
correction through $w^{(s)}_{i \vec k}$ (see text). Inset:
$\Gamma^{\uparrow,\downarrow}_1$ if both the $s$ and $d$ bands contribute in
the injected electrons. }
\end{figure}

Now the crucial point is how to single out the injected electrons that
are actually responsible for the H vibrational (de)excitation. With
this purpose, we turn to the theory developed to describe propagation
of hot electrons through a periodic lattice in
BEEM~\cite{bib:deandres01}. In its simplest version, BEEM theory uses
a semiclassical Greens function due to Koster~\cite{bib:koster54} to
link the topography of the band-structure constant energy surface at
$E_{F} \pm V$ with the spatial propagation of electrons in the
substrate periodic lattice. For Pd, it predicts the formation of
narrowly focused beams associated with the nearly flat $d$ bands,
while electrons injected in $s$ bands propagate in a semispherical,
$s$-wave-like form (see Fig.~\ref{fig:esquema}). For the experimental
samples of diluted H in Pd we expect a behavior of the $s$ and $d$
bands similar to that found in {\it bulk} Pd, since the low H
concentration has a small impact on the Pd band structure. This has
been seen in the Pd$_{16}$H supercell, that already represents well
the low-density regime in the experiment ~\footnote{The bottom of the
$d$ band is weakly contributed by H-Pd hybridization and the states
lying within 1\,eV above and below $E_F$ are unaffected by the
inclusion of H. The $d$ band upper edge spills $\sim 0.2$\,eV above
$E_F$, and bands lying at higher energies correspond to strongly
dispersive $s$-like states. The H states lie approximately 2-3\,eV
below the bottom of the conduction band.}. Therefore, electrons
injected in the strongly focussed $d$ bands are not likely to
encounter and, hence, to excite interstitial H. This mechanism can
only be driven by the $s$ bands propagating in all directions in the
crystal.
To take into account this effect, we introduce in Eq.~\ref{eq:gamma1} weights,
$w^{(s)}_{i\vec k}$, that have been obtained as the projected
$s$-orbital percentage in the whole $| \psi_i (\vec k) \rangle$
Kohn-Sham state. The calculated $\Gamma^{\uparrow,\downarrow}_1$
are plotted in Fig.~\ref{fig:gamma1}. As seen in the
inset, if both $s$ and $d$ states contributed fully,
$\Gamma^{\uparrow,\downarrow}_1$ would map the Pd DOS and the large
asymmetry around $E_F$ would cause differences
in the rates of at least 2 orders of magnitude between positive and negative $V$.
These differences are drastically reduced as soon as we realize that H
excitation is mainly caused by nearly free ($s$-like) electrons,
that have a similar dispersion above and below $E_F$.

\begin{figure}
\includegraphics[scale=0.7]{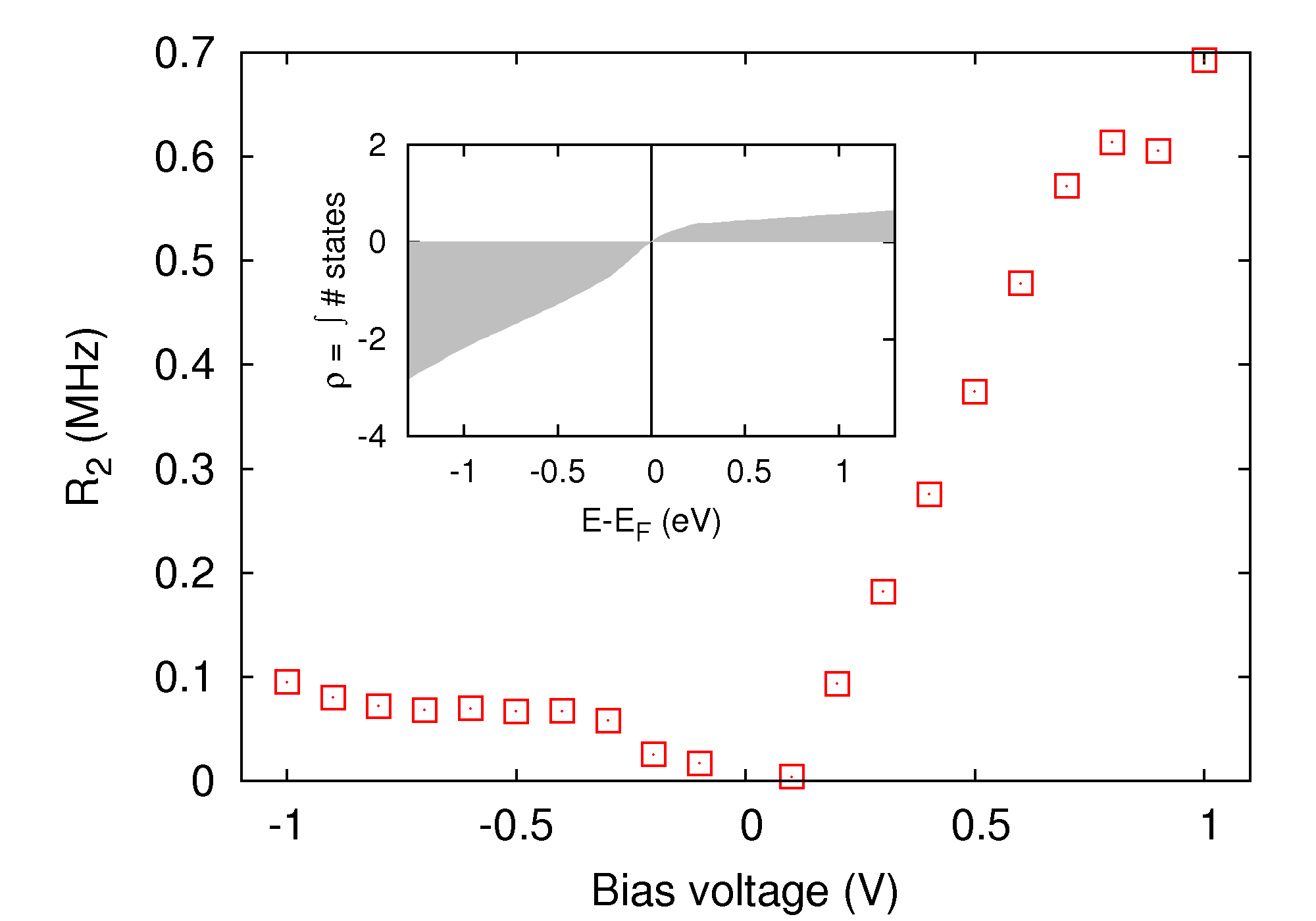}
\caption{\label{fig:RN} (Color online) H transfer rate from the third excited
vibrational state as a function of the bias voltage for $I_t = 100$\,pA. Inset:
number of available initial states per Pd atom. }
\end{figure}

Rates in Eq.~\ref{eq:gamma1} are not yet the ones appearing in
Eq.~\ref{eq:transfer}. The latter are the result of the \emph{net}
vibrational excitation caused by all the charge carriers in the
incoming current, $I_t$.  Therefore, we introduce the following
linear model for $\Gamma^{\uparrow,\downarrow}$:
\begin{equation}
\Gamma^{\uparrow} = \frac{I_t}{e \Gamma_0 \rho} \Gamma^{\uparrow}_1
\;, \hspace{0.8cm} \Gamma^{\downarrow} = \Gamma_0 + \frac{I_t}{e
\Gamma_0 \rho} \Gamma^{\downarrow}_1 \label{eq:gamma}
\end{equation}
where $\Gamma_0$ is the deexcitation rate present at zero bias
[$\Gamma_{0}$ is calculated from Eq.~\ref{eq:gamma1} simply by
removing the $w^{(s)}_{i \vec k}$ factors], and $\rho$ (see inset
in Fig.~\ref{fig:RN}) is the number of available one-electron
states per Pd atom. By inserting this factor we are assuming
that all the incoming carriers have an initial energy $\epsilon_i$
with equal probability $1/\rho$, and also that the final state is
always unoccupied. The latter is a fair assumption in the present
low-intensity regime. Considering that $s$ carriers distribute
evenly throughout the crystal and that the typical time that a
H atom spends in a given excited vibrational state is $\sim 1/
\Gamma_0 \sim 10^{-10}$\,s, the factor $\frac{I_t}{e \Gamma_0}$
represents the fraction of the current that (de)excites the H atom.
The weak inelastic limit ($\Gamma^{\downarrow} \simeq \Gamma_0 \gg
\Gamma^{\uparrow}$) holds well for the experimental intensities
$I_t = 1-150$\,pA.
In particular, for $V=\pm 0.7$\,V and $I_t=150$\,pA the
effect of the injected hot electrons can be understood as if the interstitial
H was effectively surrounded by a thermal bath at $T_{\mathrm{eff}} \simeq 80$\,K
according to the expression for a Botzmann distribution of level populations
$k_B T_{\mathrm{eff}} = \hbar \omega / \mathrm{ln} (\Gamma^{\downarrow} / \Gamma^{\uparrow})$
\cite{bib:gao97}.

Quantum tunneling becomes important in the diffusion of nonmassive
elements, like H, specially at low temperatures.  We introduce it
in the model via a semiclassical (WKB) transmission rate through the
barrier, $T(E_n)=e^{-2\gamma(E_n)}$.  To a very good approximation,
a triangular-shaped barrier can be used instead of the upper part of
the pathway in Fig.~\ref{fig:barrier}. In this case, $\gamma(E_n) =
\frac{2}{3}\big[ \frac{2m_H}{\hbar^2}(E_b-E_n) \big]^{1/2} d(E_n)$,
where $d(E_n)$ is the barrier width seen by H at the $n$-th vibrational
level.  We obtain $T(E_2)=5.1\times10^{-3}$ and $T(E_3)=0.24$.
H transfer will occur as a result of climbing a
$n$-step ladder and tunneling from this energy level with a combined
transfer rate $R_{n}T(E_{n})$. Since $R_2T(E_2) > R_3T(E_3) \gg R_4$,
at the experimental currents of 1-150~pA, $n=2$ is
the optimum level for the combined mechanism.  Figure~\ref{fig:RN}
shows $R_2$ for $I_t=100$\,pA.
Values at different $I_t$ are
estimated by noting that $R_n$ scales as $I_{t}^{n}$ in the weak
inelastic limit. This scaling also shows that H transfer from the
second and third vibrational states becomes comparable at larger
currents ($I_t > 1$\,nA). Importantly,
the similar rates obtained with positive and negative $V$
are consistent with the experimental insensitivity to the bias
polarity.
Furthermore, we observe in Fig.~\ref{fig:RN} a
tendency to yield larger rates at $V\gtrsim 0.2$\,V, consistent
with the slightly wider stripes reported by Sykes {\it et al.}
at $V>0$.
This effect is a direct consequence of the distinct
excitation dynamics of holes and electrons. More precisely, if the
focusing of $d$ carriers had not been considered, the
huge asymmetry in $\Gamma^{\uparrow,\downarrow}_1$ (see inset in
Fig.~\ref{fig:gamma1}) would have favored negative voltages over positive ones in $R_n$
(e. g. $> 1$ order of magnitude in $R_2$).

In conclusion, we unravel the mechanism behind H diffusion in bulk Pd assisted
by inelastic interaction with STM-injected hot electrons. Experiments reported
for this system exhibit comparable H transfer for $V>0$ and $V<0$. This is a
counter-intuitive result considering the strong difference between electron and
hole band structures in Pd.  We find that both observations are reconciled by
the fact that carriers from Pd $d$ bands do not assist H diffusion, as their
propagation is restricted to narrowly focused cones. The H
vibrational excitation is only driven by the remaining $s$ carriers,
more symmetrically distributed around $E_F$. Moreover, we find that at the
experimental currents of 1-150~pA quantum tunneling is
crucial to explain H diffusion (without it, the rates would be at least 2
orders of magnitude lower).
Our model allows us to quantify the dependence of the diffusion rates
on the experimental $I_t$ and $V$ at low temperature.
Since this mechanism is based in general concepts of solid-state theory i.e.,
the electronic band structure in a periodic potential, the propagation according to
related Green functions, the e-ph interaction and tunneling of atoms
at low temperatures, we believe our conclusions should apply to other transition metals.

\begin{acknowledgments}
We thank V. Chis and R. D\'{\i}ez Mui\~no for stimulating discussions.
M.B.-R. acknowledges financial support from the Gipuzkoako Foru Aldundia and
the European Commission Project No. FP7-PEOPLE-2010-RG276921.
This work is also supported by the
Gobierno Vasco-UPV/EHU Project No. IT-366-07, and the Spanish MICINN
Projects No. FIS2010-19609-C02-02 and No. MAT2011-26534.
Computational resources were provided by the DIPC computing center.
\end{acknowledgments}

\bibliography{hpdC-shorten}

\end{document}